\documentclass[amssymb,aps,showpacs,nofootinbib,twocolumn]{revtex4}
\usepackage[]{graphicx}
\usepackage[]{amsmath}
\usepackage{hyperref}

\def\ket#1{| #1 \rangle}
\def\bra#1{\langle #1 |}

\def\kb#1#2{| #1 \rangle\!\langle #2 |}
\def\II{1\!\mathrm{l}}
\def\cH{\mathcal{H}}
\def\cX{\mathcal{X}}
\def\bL{{\bf L}}
\def\bR{{\bf R}}
\def\el{{\boldsymbol{\ell}}}

\begin{document}

\title{Macroscopic observables}
\author{David Poulin}
\email{dpoulin@iqc.ca}
\affiliation{
Perimeter Institute for Theoretical Physics, Waterloo, ON, Canada. \\ 
Institute for Quantum Computing, University of Waterloo, Waterloo, ON, Canada.\\
Department of Physics, University of Waterloo, Waterloo, ON, Canada.}

\date{\today}

\begin{abstract}
We study macroscopic observables defined as the total value of a physical quantity over a collection of quantum systems. We show that previous results obtained for {\em infinite} ensemble of identically prepared systems lead to incorrect conclusions for finite ensembles.  In particular, exact measurement of a macroscopic observable significantly disturbs the state of any finite ensemble. However, we show how this disturbance can be made arbitrarily small when the measurement are of finite accuracy. We demonstrate a tradeoff between state disturbance and measurement coarseness as a function of the size of the ensemble. Using this tradeoff, we show that the histories generated by any sequence of finite accuracy macroscopic measurements always generate a consistent family in the absence of large scale entanglement, for sufficiently large ensembles. Hence, macroscopic observables behave ``classically" provided that their accuracy is coarser than the quantum correlation length-scale of the system. The role of these observable is also discussed in the context of NMR quantum information processing and bulk ensemble quantum state tomography. 
\end{abstract}

\pacs{03.65.Ta, 03.67.-a}

\maketitle

\section{Overview}\label{intro}

Macroscopic observables correspond to physical quantities which are accessible to our senses. Since the physical scale of individual quanta is generally tiny, macroscopic observables arise when a collection of quantum systems are measured jointly. Formally, they can be described by {\em type projectors}, which reveal information about the average population of single-particle states. For example, the total magnetization of an ensemble of spin-$\frac 12$ particles provides some information about the relative occupation number of the $up$ and $down$ states. We will derive several general properties of these measurement and discuss how they lead to the emergence of a quasiclassical domain in the absence of large scale entanglement. 

The effect of macroscopic observations on infinite ensemble of identically prepared quantum systems has been studied in various contexts \cite{Finkelstein63,Hartle68,Graham73,FGG89}. The main conclusion of these studies is that the state $\ket\psi^{\otimes N}$ describing such an ensemble is an eigenstate of type projectors when $N=\infty$. However, for finite ensembles, things change dramatically. The measurement of a macroscopic observable induces a disturbance which {\em increases} as the size of the ensemble grows, in apparent contradiction with the infinite-copy result. This discrepancy follows from the ambiguous extension of finite-copy considerations to the nonseparable Hilbert space of an infinite-copy ensemble \cite{CS2004}. In this article, we show how the essence of the infinite-copy result can be recovered by ``smoothing" the type projectors into coarse-grained positive operator valued measurement (POVM) (essentially going from the strong to the weak law of large numbers).  

The paper is organized as follows. The central mathematical objects of the present study are defined in Section~\ref{definitions}. We first summarize the method of type and define type projectors. These are projectors on the degenerated eigensubspaces of macroscopic observables of the form $A_N = \sum_{k=1}^N a_{(k)}$, where $a_{(k)}$ acts on the $k$th system of the ensemble. Using the theory of generalized measurements, we also define coarse-grained POVMs corresponding to finite accuracy estimation of a macroscopic observable. 

Section~\ref{main} contains the core mathematical analysis of our study. We first recapitulate the well known facts about type projectors acting on infinite ensemble  and show how they dramatically break down for finite ensemble.  Then, we show how the result is approximately recovered when the measurements are of finite accuracy, and study the general tradeoff between measurement coarseness and state disturbance --- measured in terms of fidelity --- as a function of the size of the ensemble. In short, we demonstrate that a measurement of coarseness $\sigma \gg 1/\sqrt N$ leaves the systems essentially unchanged, i.e. the fidelity $F$ between the pre- and post-measurement state of the ensemble satisfies $1-F \propto \frac{\ln(N\sigma^2)}{N\sigma^2}$. 

Section~\ref{exchangeability} is a discussion of the de Finetti representation theorem which provides a wide class of state --- exchangeable states --- for the study of macroscopic observables. Exchangeable states have recently been employed for the discussion of quantum state tomography based on single-system measurements followed by Bayesian update~\cite{CFS2002}. We will show how macroscopic observables offer an alternative perspective on quantum tomography. Moreover, this approach offers interesting applications for quantum information theory~\cite{HM2002,BHL2004} and is a more accurate description of experimental spectroscopy-based implementations of tomography, e.g. as achieved in Ref.~\cite{VFPKLC2001}.

Macroscopic observables also provide an explanation for the emergence of the classical world we perceive from the underlying quantum theory. Indeed, we demonstrate in section~\ref{classicality} that in the absence of large-scale entanglement,  one of the main characteristic of the classical domain --- namely the commutativity of classical observables --- follows naturally from the general properties of coarse grained type POVMs. This is done using the consistent histories formalism which we briefly summarize. We demonstrate that the histories generated by {\em any} sequence of macroscopic observables of accuracy $\sigma \gg \sqrt{\xi/N}$ are consistent, where $\xi$ is the quantum correlation length-scale of the system. This generalizes some ideas introduced by Halliwell~\cite{Halliwell} on how to achieve classicality in closed quantum systems.  

Finally, Section~\ref{NMR} discusses the role of macroscopic observables in  NMR quantum information processing. In this context, macroscopic observables are used to extract the output of the computation, but also, since the measurement device can not be ``turned off'' --- i.e. the state of the processor can always be read-off from the spectrometer --- they constantly perturb the computation.  Following the results of Section~\ref{main} and a measurement model introduced in Ref.~\cite{LS2000}, we show that the measurements used in NMR can in principle be sufficiently precise to extract useful information about the computation but yet so coarse grained that they induce a negligible perturbation. However, as we will demonstrate, NMR measurements may not follow our optimal measurement coarseness-state disturbance tradeoff when performed at room temperature; caution is advised when applying our conclusions. 

\section{Definitions}\label{definitions}

This section contains all the mathematical definitions required for our study. For sake of clarity, we adopt the vocabulary of NMR. Our general setting consists of an ensemble of $N$ quantum systems of the same nature. Therefore, we shall refer to individual systems of an ensemble as {\em molecules} and to the ensemble of $N$ molecules itself as the {\em sample}. Thus, the word ``molecule" should not be taken literally in what follows; it could be any elementary constituent of a larger system.  

\subsection{Method of types}\label{types}

The method of type is a very powerful statistical tool with applications ranging from large deviation theory, universal coding, and hypothesis testing. We will only scratch the surface of this theory here, more details and applications can be found in \cite{CT1991} for instance.

Let $X = x_{j_1}x_{j_2}\ldots x_{j_N} \in \mathcal X^N$ be a string of $N$ letters drawn from a $d$-letter alphabet $\cX = \{x_1,x_2,\ldots x_d\}$. The {\em type} (or empirical probability distribution) of $X$ is a vector of positive numbers summing to one defined by
\begin{equation*}
{\bf L}(X) = \Big(L_1(X),L_2(X),\ldots, L_d(X)\Big),
\end{equation*}
where $L_j(X)$ is the relative frequency of the letter $x_j$ in the string $X$
\begin{equation*}
L_j(X) = \frac 1N \sum_{k=1}^N \delta_{j,j_k};
\end{equation*} 
it is simply the number of occurrences of the letter $x_j$ in $X$, divided by the length of $X$.
For example, if $\cX = \{a,b,c\}$ and $N = 4$, then $\mathbf{L}(cbaa) = (\frac 12,\frac 14,\frac 14)$. 
We also define a type class $T$ to be the set of strings of a given type:
\begin{equation*}
T[{\bf L}] = \{X \in \cX^N: {\bf L}(X) = {\bf L}\}.
\end{equation*}
For example, using the same alphabet as above, we have $T[(\frac 14,0,\frac 34)] = \{accc,cacc,ccac,ccca\}$. The class $T[{\bf L}]$ can be generated by applying all permutations to any single string of type ${\bf L}$. Hence, the number of elements in $T[{\bf L}]$ is given by the multinomial coefficient
\begin{eqnarray*}
|T[{\bf L}]| &=& \binom{N}{NL_1,NL_2,\ldots,NL_d} \\
&=& \frac{N!}{(NL_1)!,(NL_2)!,\dots,(NL_d)!}
\end{eqnarray*}

Let ${\bf R} = (R_1,R_2,\ldots,R_d)$ be a probability distribution over $\cX$. The probability of the string of outputs $X = x_{j_1}\ldots x_{j_N}$ of $N$ letters, each drawn independently according to the distribution $\bf R$, is $P(X) = R_{j_1}R_{j_2}\ldots R_{j_N}$. This can also be written as  
\begin{equation*}
P(X) = R_1^{NL_1(X)}R_2^{NL_2(X)}\ldots R_d^{NL_d(X)},
\end{equation*}
so given a fixed distribution $\bf R$, the probability of a string $X \in \mathcal X^N$ depends only on its type. Intuitively, the type of the observed outcome $X$ is very likely to be close to the probability distribution of the random variable, i.e. $L_j(X) \approx R_j$, as $N$ increases. This is the substance of the {\em typical sequence theorem}~\cite{CT1991}:
\begin{equation}
P(\|{\bf L} - {\bf R}\|_1^2 > \epsilon) \leq e^{-N(\epsilon/2 - d\frac{\ln(N+1)}{N})} \approx e^{-N\epsilon/2}
\label{eq:tst}
\end{equation}
where the ``difference'' between the type ${\bf L}$ and the probability distribution ${\bf R}$ is quantified by the variational distance ($L_1$-norm)
\begin{equation*}
\|{\bf L} - {\bf R}\|_1 = \sum_j |L_j(X) - R_j|.
\end{equation*}
The typical sequence theorem takes on various forms. It can be formulated in a stronger version using the relative entropy, which is an upper bound to the variational distance. Nevertheless, for our considerations, this simple version will be sufficient. 

\subsection{Macroscopic observable}\label{measurements}

Using this notation, we now formally define macroscopic observables. Consider a Hermitian operator (i.e. {\em observable}) $a$ acting on the $d$-dimensional Hilbert space of a single molecule $\cH_m$. Let $\{\ket{x_1}, \ket{x_2}, \ldots \ket{x_d}\}$ and $\{\alpha_1,\alpha_2,\ldots,\alpha_d\}$ denote its eigenvectors and eigenvalues: $a\ket{x_j} = \alpha_j\ket{x_j}$. We will assume that $a$ is {\em non-degenerate}, generalization is straightforward. The macroscopic observable $A_N$ corresponds to the sum of observable $a$ over all the $N$ molecules of the sample,
\begin{equation}
A_N = \sum_{k=1}^N a_{(k)},
\label{eq:macroscopic}
\end{equation}
where $a_{(k)}$ is the operator $a$ acting on the $k$th molecule, 
\begin{equation*}
a_{(k)} = \underbrace{\II\otimes\ldots\otimes\II}_{k-1}\otimes \ a\otimes\underbrace{\II\otimes\ldots\otimes\II}_{N-k}.
\end{equation*}
The operator $A_N$ acts on the joint Hilbert space of the $N$ molecules $\cH_s = \cH_m^{\otimes N}$ --- the Hilbert space of the sample --- which has dimension $d^N$. We use the standard abbreviation $\ket X = \ket{x_{j_1}}\otimes\ket{x_{j_2}}\otimes\ldots\otimes\ket{x_{j_N}}$ for each string  $X \in \cX^N$. Clearly, the states $\ket X$ form an orthonormal basis for $\cH_s$. Moreover, they are eigenstates of the macroscopic observable $A_N$:
\begin{eqnarray*}
A_N\ket X &=& A_N\ket{x_{j_1}}\otimes\ket{x_{j_2}}\otimes\ldots\otimes\ket{x_{j_N}} \\
&=& \sum_{k=1}^N a_{(k)} \ket{x_{j_1}}\otimes\ket{x_{j_2}}\otimes\ldots\otimes\ket{x_{j_N}} \\
&=& \sum_{k=1}^N \alpha_{j_k} \ket{x_{j_1}}\otimes\ket{x_{j_2}}\otimes\ldots\otimes\ket{x_{j_N}} \\
&=& \sum_{k=1}^N \alpha_{j_k} \ket X = \left(\sum_{j=1}^d N L_j(X)\alpha_j\right) \ket X.
\end{eqnarray*}
Thus, we see that the eigenvalue associated to a basis state $\ket X$ depends only on its type ${\bf L}(X)$. As a consequence, the degenerated eigensubspaces of $A_N$ are those subspaces spanned by the vectors $\ket X$ belonging to the same type class. 

This brings us to the definition of {\em type measurements} which are von Neumann  measurement composed of the projection operators on the subspaces of a given type:
\begin{equation}
Q_{\bf L}^{(N)} = \sum_{ X \in T[{\bf L}] }\kb XX.
\label{eq:proj_pop}
\end{equation}
Each of these {\em type projectors} is labeled by a vector of $d$ positive numbers $L_j$ which correspond to the type ${\bf L}(X)$ of the basis vectors $\ket{X}$ spanning the subspace. Obviously, the projectors $ Q_{\bf L}^{(N)}$ depend on the choice of basis $\ket{x_j}$ over $\cH_m$, i.e. on the eigenvectors of the observable $a$, so we could explicitly note $Q_{\bf L}^{(N,a)}$. Moreover, we would like to stress that the spectral projectors $Q_{\bf L}^{(N,a)}$ and $Q_{\bf L'}^{(N,b)}$ associated to two {\em distinct} macroscopic observables $A_N = \sum_k a_{(k)}$ and $B_N = \sum_kb_{(k)}$ {\em do not commute}, unless the underlying single-molecule observables $a$ and $b$ happen to commute. To avoid cumbersome notation, we will only use an extra superscript when necessary (c.f. Section~\ref{tomography}). For the time being, we will consider a fixed arbitrary macroscopic observable $A_N$. In this case, it is straightforward to verify that the type projectors are mutually orthogonal and that they sum to the identity
\begin{equation}
Q_{\bf L}^{(N)}Q_{{\bf L}'}^{(N)} = \delta_{{\bf L},{\bf L}'}Q_{\bf L}^{(N)}
\ ,\ \ \ 
\sum_{\bf L} Q_{\bf L}^{(N)} = \II .
\label{eq:orthogonal}
\end{equation}
In words, these projectors correspond to the exact measurement of the population of the levels $\ket{x_j}$ over an ensemble of $N$ molecules, without distinguishing between the molecules of the sample. The type projectors $Q_{\bf L}^{(N)}$ allows us to express the operator $A_N$ in a simple form:
\begin{equation}
A_N = \sum_{\bf L} A_{\bf L} Q_{\bf L}^{(N)}
\label{eq:spectral}
\end{equation}
where we have defined $A_{\bf L} = \sum_{j=1}^d N L_j\alpha_j$. Similarly, any macroscopic observable of the form Eq.(\ref{eq:macroscopic}) has a spectral decomposition involving only type projectors, as in Eq.(\ref{eq:spectral}). Hence, following textbook quantum mechanics, when measuring a macroscopic observable --- or measuring the ``expectation value" of a physical observable over a macroscopic sample ---, one is really performing a projective von Neumann measurement composed of type projectors. 

These type projectors have been studied under many different forms \cite{Finkelstein63,Hartle68,Graham73} and take on many different names. Among other formulation are the frequency operators. Recall that $L_j(X)$ is the relative frequency of the symbol $x_j$ in the string $X$. We can define a {\em frequency operator} 
\begin{equation*}
F^{(N)}_j = \sum_X L_j(X) \kb XX.
\end{equation*}
This operator is a macroscopic physical observable whose eigenvalues are $f_j = 0,\frac 1N, \frac 2N, \ldots,1$. Indeed, $F_j^{(N)}$ takes on the form of Eq.~(\ref{eq:macroscopic}) by setting the single-molecule observable $a$ to $\frac 1N \kb{x_j}{x_j}$. Following textbook quantum mechanics, when the measurement associated to $F_j^{(N)}$ is performed and eigenvalue $f_j$ is observed, the state of the system gets collapsed to the subspace spanned by the states $\ket X$ for which $L_j(X) = f_j$. Hence, the eigenvalue $f_j$ indicates the relative population of the single-molecule state $\ket{x_j}$ in the sample of $N$ molecules.  

The above construction yields $d$ {\em commuting} physical observable $\{F_j^{(N)}\}_{j=1,\ldots, d}$, one for each single-molecule state $\{\ket{x_j}\}_{j=1,\ldots,d}$. Regrouping these observable into a $d$-component observable yields
\begin{eqnarray}
{\bf F^{(N)}} &=& (F^{(N)}_1,F^{(N)}_2,\ldots,F^{(N)}_d) \label{eq:frequency}\\
&=& \sum_{\bf L}{\bf L}Q_{\bf L}^{(N)}, \nonumber
\end{eqnarray}
which takes on the form of Eq.(\ref{eq:spectral}), with a $d$-component eigenvalue $A_{\bf L} = \bL$. The value of any macroscopic observable of the form Eq.(\ref{eq:spectral}) can be deduced straightforwardly from the value of ${\bf F}^{(N)}$. Hence, a great deal of attention has been  focused on the macroscopic observable ${\bf F}^{(N)}$ without loss of generality. 

We illustrate macroscopic observables for a sample of $N$ spin-$\frac 12$ particle. We choose the basis $\ket{x_1} = \ket{\!\!\uparrow}$ and  $\ket{x_2} = \ket{\!\!\downarrow}$  corresponding, respectively, to $+\frac 12$ and $-\frac 12$ units of magnetization in the $z$ direction: 
\begin{equation*}
\sigma^z \ket{x_1} = \frac 12 \ket{x_1}\ \mathrm{and}\  
\sigma^z\ket{x_2} = -\frac 12 \ket{x_2}.
\end{equation*}
We can use a single positive number $L \in \{0,\frac 1N,\frac 2N,\ldots, 1\}$ to label a type of a binary string $X$, which corresponds to the fraction of $x_1$'s (or {\em spin up}'s) in $X$. Hence, a type $L$ is a shorthand for ${\bf L} = (L,1-L)$. The bulk (or total) magnetization of the sample is equal to the sum of the magnetization of each molecules: the corresponding operator is therefore $M^z = \sum_{k=1}^N \sigma^z_{(k)}$, where $\sigma^z_{(k)}$ is the ``$z$" Pauli operator acting on the $k$th molecule. When the sample is in a state of a definitive type, its bulk magnetization is equal to $\frac 12 N(L_1 - L_2) = \frac 12 N(2L - 1)$, which is simply the number of spins pointing up minus the number of spins pointing down, times $\frac 12$. Hence, the observable corresponding to the bulk magnetization can be written as
\begin{equation*}
M^z = N\frac 12\sum_L (2L-1)Q_L^{(N)}
\end{equation*}
where the sum is over all types. The type projectors $Q_L^{(N)}$ are projectors on the degenerated eigensubspaces of the bulk magnetization operator. Clearly, an exact measurement of the magnetization $M^z$ would reveal the type of the state of the sample, i.e. the relative frequency of {\em up} and {\em down} spins. 

\subsection{Coarse grained macroscopic POVMs}

We will now present how finite accuracy macroscopic observables can be expressed in terms of type projectors. Before we do so, we briefly recall some basic concepts of the theory of generalized measurement. Generalized measurements (POVMs) are described by a set positive operators $E_j$ summing to identity. The generalized Born rule for the probability of getting outcome $E_j$ given initial state $\rho$ is the same as for von Neumann measurements
\begin{equation}
P(E_j|\rho) = Tr\{E_j\rho\}.
\label{eq:general_prob}
\end{equation}
After the measurement outcome $E_j$ is observed, the state of the system gets updated to
\begin{equation}
\rho \xrightarrow{j} \rho_{|j} =  \frac{\sum_i A_{ji}^\dagger \rho A_{ji}}{P(E_j|\rho)},
\label{eq:general_update}
\end{equation}
where the {\em Kraus operators} $A_{ji}$ can be any set of operators satisfying $\sum_i A_{ji}A_{ji}^\dagger = E_j$. Here, we will often consider {\em ideal} quantum measurements where the disturbance inflicted to the system is minimal. This restriction is necessary if we want to study the optimal tradeoff between information gathering and state disturbance. To each measurement outcome $E_j$ of an ideal measurement is associated a {\em single} Kraus operator $A_{j0} = \sqrt{E_j}$. In this case, the state update rule Eq.(\ref{eq:general_update}) simplifies to
\begin{equation}
\rho \xrightarrow{j}  \rho_{|j} =  \frac{\sqrt{E_j}^\dagger \rho \sqrt{E_j}}{P(E_j|\rho)},
\label{eq:general_proj}
\end{equation}
which reduced to the regular state update rule when $E_j$ are projection operators. Hence, von Neumann measurements are POVMs with an extra orthogonality constraint. Any generalized measurement can be realized physically by coupling the system of interest to a larger system and performing a von Neumann measurement on the larger system; an example of such a physical construction will be presented in Section~\ref{NMR}. Similarly, any such ``indirect'' measurement corresponds to a POVM. Hence, POVMs do not add anything extra to plain textbook quantum mechanics, beside conciseness. 

Continuing with our example, {\em finite accuracy} measurement of the bulk magnetization of a sample of $N$ spin-$\frac 12$ molecules can be described in terms of {\em coarse-grained} type operators $\tilde{Q}_\ell^{(N)}$. When the state of the sample is of a definite type $L$, the {\em observed} value of the bulk magnetization will not necessarily be equal to $\frac 12 N(2L-1)$ but, due to the uncertainty of the measurement apparatus, may take different values $\frac 12 N(2\ell-1)$, with respective probabilities $q_L(\ell)$. The function $q_L(\ell)$ should be centered around $L$ and have a certain width $\sigma$ corresponding to the coarseness of the measurement. 

Hence, the coarse-grained type measurements can be defined by ``smoothing" the exact type projectors: 
\begin{equation}
\tilde{Q}_{\boldsymbol{\ell}}^{(N)} = \sum_{{\bf L}} \sqrt{q_{\bf L}({\el})} Q_{{\bf L}}^{(N)}
\label{eq:cg_proj}
\end{equation}
where $q_{\bf L}({\el})$ is some probability distribution over $\el$ centered roughly at ${\bf L}$ and has the interpretation given above. In principle, $\el$ could be any real $d$-dimensional vector, as it contains statistical fluctuations. For example, $q_{\bf L}({\el})$ could be a $d$-dimensional Gaussian 
\begin{equation}
q_{\bf L}({\el}) = \left(\frac{1}{2\pi\sigma^2}\right)^{\frac d2}
\exp\left\{-\frac{\|{\el}-{\bf L}\|^2_2}{2\sigma^2}\right\}
\label{ex_prob}
\end{equation}
which is properly normalized $\int q_{\bf L}({\el}) d\el = 1$ and where the $L_2$-norm is $\|\el-\bL\|_2^2 = \sum_j (\ell_j-L_j)^2$. The operators $E_{\el} =  \tilde{Q}_{\el}^{(N)}\tilde{Q}_{\el}^{(N)\dagger}$ form a POVM (with a continuous outcome) since they are all positive operators and satisfy
\begin{equation}
\int E_\el d\el = \int \tilde{Q}_{\el}^{(N)}\tilde{Q}_{\el}^{(N)\dagger} d\el = \II .
\label{eq:id_POVM}
\end{equation}
These coarse-grained type operators describe a situation where our measurement apparatus is not sufficiently precise to measure the exact population of each level, but rather provides an estimation of it within a finite accuracy $\sigma$.

We have assumed that the measurement outcome $\el$ takes on a continuous spectrum. However, several measurement apparatus, like those equipped with a numerical output display, have a discrete spectrum of outcomes. This can be taken into account by choosing a smoothing function
\begin{equation*}
q_\bL(\el) = \sum_{\el_j} \delta(\el-\el_j) f_j(\bL)
\end{equation*}
where $\{\el_j\}$ is the set of possible outcomes. Thus, we will henceforth consider the more general continuous case, but all our analysis carries through for discrete measurement outcomes by performing the above substitution. 

\section{Type measurement on identically prepared systems}\label{main}

Type projectors were first studied by Finkelstein~\cite{Finkelstein63}, Hartle~\cite{Hartle68}, and Graham~\cite{Graham73} as part of  discussions on the interpretation of probabilities in quantum theory. The main characteristic of type projectors identified by these authors can be summarized as follows. Let $\ket \psi = \sum_j\beta_j \ket{x_j}$ be an arbitrary pure state of a $d$-level molecule, with associated density matrix $\nu = \kb \psi\psi$. Consider a sample of $N$ identically prepared molecules, such that the state of the sample is $\ket{\Psi_N} = \ket\psi^{\otimes N}$. Upon measurement of the type of the sample, we expect a result close to the probability distribution ${\bf R} = (\bra{x_1}\nu\ket{x_1},\ldots,\bra{x_d}\nu\ket{x_d}) = (|\beta_1|^2,\ldots,|\beta_d|^2)$. Indeed, it follows from the strong law of large numbers that
\begin{equation}
\lim_{N \rightarrow \infty} 
\Big| F^{(N)}_j \ket{\Psi^N} - |\beta_j|^2 \ket{\Psi^N}\Big|^2 = 0,
\label{eq:hartle}
\end{equation}
where $F^{(N)}_j$ is the $j$th component of the frequency operator defined at Eq.(\ref{eq:frequency}). In other word, ${\bf F}^{(N)}\ket{\Psi_N} = {\bf R}\ket{\Psi_N}$ with probability one in the limit of infinite $N$. This led Hartle to the conclusion that an infinite number of identically prepared molecules are in an eigenstate $\ket{\Psi_\infty}$ of the frequency operator ${\bf F}^{(\infty)}$ with eigenvalue $\bR$. Finkelstein, on the other hand, concluded from Eq.(\ref{eq:hartle}) that for finite $N$, $\ket{\Psi_N}$ is ``close" to an eigenstate of ${\bf F}^{(N)}$ with eigenvalue ${\bf R}$. Thus, a measurement of the frequency operator reveals the probabilities $R_j = \bra{x_j}\nu\ket{x_j}$,  in the standard Copenhagen sense, of observing a single molecule of the sample in the state $\ket{x_j}$.

However, the conclusions reported above can be quite misleading. There are really two distinct issues here. The first one concerns the validity of the argument as a derivation of Born's rule to assign probabilities in quantum theory. The main complication comes from the definition of $\bf F^{(\infty)}$ as the limit of a finite operator. This limit does not uniquely defined the operator on the {\em non-separable} Hilbert space $\cH_m\otimes\cH_m\otimes\ldots$ of the infinite sample: specifying the action of $\bf F^{(\infty)}$ on all states of the form $\ket{x_1}\otimes\ket{x_2}\otimes\ldots$ is not enough to define it. This was realized in \cite{FGG89} where an alternative derivation of the probability rule was presented. Nevertheless, the proposed solution is still not satisfactory as it relies itself on probability theory. An up-to-date and rather critical discussion of the status of the frequency operator and the related programs can be found in a recent paper of Caves and Schack \cite{CS2004}. We will not address these issues any further and do not claim to offer an alternative derivation of Born's rule.

The second difficulty which is directly relevant to the present study concerns state disturbance. When a system is prepared in an eigenstate of a physical observable, the act of measurement does not disturb it. While Eq.(\ref{eq:hartle}) does not grant this for any finite $N$, one naturally expects that, as $N$ grows, the disturbance caused by the measurement should decrease and eventually become negligible for all practical purpose. 

In what follows, we will show that {\em the measurement of macroscopic observables induces an importance disturbance to the state of the sample}. In fact, {\em this disturbance increases as the size $N$ of the sample grows}. This is in apparent contradiction with the conclusion that one might intuitively draw from Eq.(\ref{eq:hartle}) by extending it to finite $N$. However, we will show how the above conclusion can be recovered when the measurement of macroscopic observables are of finite accuracy: {\em sufficiently coarse grained type measurements induce a negligible disturbance to the state of the sample}. We are interested in the tradeoff between measurement accuracy and state disturbance. 

\subsection{State disturbance}

The state of the sample can be rearranged as follows
\begin{eqnarray}
\ket{\Psi_N} &=& \left(\sum_{j=1}^d \beta_j \ket{x_j}\right)^{\otimes N} \nonumber \\
&=& \sum_{X \in \mathcal{X}^N} \left(\prod_{j=1}^d \beta_j^{NL_j(X)}\right) \ket X \nonumber \\
&=& \sum_\bL  \left[\prod_{j=1}^d \beta_j^{NL_j}\right] \sum_{X \in T[\bL]} \ket X \nonumber \\
&=& \sum_\bL  \left(\prod_{j=1}^d \beta_j^{NL_j}\right) \sqrt{|T[\bL]|} \ket \bL 
\label{eq:state_copy}
\end{eqnarray}
where we have defined the normalized state
\begin{equation*}
\ket \bL = \frac{1}{\sqrt{|T[\bL]|}}\sum_{X\in T[\bL]} \ket{X}
\end{equation*}
and $|T[\bL]| = \binom{N}{NL_1,\ldots,NL_d}$ denotes the cardinality of the type class $T[\bL]$. The density operator associated to this state will be denoted $\rho_N = \kb{\Psi_N}{\Psi_N} = \nu^{\otimes N}$.  

Upon measurement of the coarse-grained operators of Eq.(\ref{eq:cg_proj}), the probability of observing an outcome within an infinitesimal volume range $d\el$ of $\el$ is $P(\tilde Q_\el^{(N)}|\rho_N) d\el$ where (see Eq.(\ref{eq:general_prob}))
\begin{eqnarray}
P(\tilde Q_\el^{(N)}|\rho_N) &=& Tr\left\{\tilde Q^{(N)}_\el \rho_N \tilde Q^{(N)}_\el\right\} \label{eq:prob_type}\\
&=& \sum_{\bL,\bL'} \sqrt{q_\bL(\el)q_{\bL'}(\el)} Tr\{Q^{(N)}_{\bL}\rho_NQ^{(N)}_{\bL'}\} \nonumber\\
&=& \sum_{\bL} q_\bL(\el) \bra{\Psi_N}Q^{(N)}_{\bL}\ket{\Psi_N}\nonumber\\
&=& \sum_\bL q_\bL(\el) m(\bL,\bR)
\label{eq:probability_ell}
\end{eqnarray}
and $m(\bL,\bR)$ denotes the multinomial distribution $\binom{N}{NL_1,\ldots,NL_d} \prod_j R_j^{NL_j}$. 
Following Eq.(\ref{eq:general_proj}), the {\em conditional post-measurement state} of the ensemble given measurement outcome $\el$ is
\begin{eqnarray}
\rho_{N|\el} &=&  \frac{\tilde Q^{(N)}_\el \rho_N \tilde Q^{(N)}_\el}{P(\tilde Q_\el^{(N)}|\rho_N)} \nonumber \\
&=& \sum_{\bL} \sum_{\bL'} \prod_{j,j'}\beta_j^{NL_j}\beta_{j'}^{*NL'_{j'}} \nonumber\\
 && \sqrt{q_\bL(\el)q_{\bL'}(\el)} \sqrt{|T[\bL]|\cdot|T[\bL']|} \ \kb{\bL}{\bL'}. 
 \label{eq:ave_rho_post}
\end{eqnarray}
The post-measurement state is obtained by averaging the conditional post-measurement states over all measurements outcomes
\begin{eqnarray}
\rho_N' &=& \int P(\tilde Q_\el^{(N)}|\rho_N) \rho_{N|\el} d\el \nonumber \\
&=& \int \tilde Q^{(N)}_\el \kb\Psi\Psi \tilde Q^{(N)}_\el \nonumber \\
&=& \sum_{\bL} \sum_{\bL'} \prod_{j,j'}\beta_j^{NL_j}\beta_{j'}^{*NL'_{j'}} \nonumber\\
 && \times G(\bL,\bL') \sqrt{|T[\bL]|\cdot|T[\bL']|} \ \kb{\bL}{\bL'}
\label{eq:rho_post}
\end{eqnarray}
where we have defined the {\em decoherence kernel}
\begin{equation}
G(\bL,\bL') = \int \sqrt{q_\bL(\el)q_{\bL'}(\el)} d\el.
\label{eq:kernel}
\end{equation}
Notice that setting $G(\bL,\bL') = 1$ in Eq.(\ref{eq:rho_post}) would yield a density matrix $\rho_N'$ identical to $\rho_N$.
Finally, the post-measurement state of a single molecule of the sample is obtained by taking a partial trace over $N-1$ molecules $\rho_{1}' = Tr_{N-1}\{\rho_{N}'\}$, and similarly for the {\em conditional} post measurement state $\rho_{1|\ell} = Tr_{N-1}\{\rho_{N|\ell}\}$.

The disturbance caused by the measurement is evaluated with the {\em fidelity} between the pre- and post-measurement state. A fidelity of $1$ indicates that the two states are identical --- i.e. the measurement did not cause disturbance --- while a fidelity $0$ indicates maximal disturbance. The fidelity between two states $\rho$ and $\nu$ is 
\begin{equation}
F(\rho,\nu) = \left(Tr\left\{\sqrt{\rho^{\frac 12}\nu\rho^{\frac 12}}\right\}\right)^2.
\end{equation}
 If one of the state is pure, say $\nu = \kb \phi\phi$, this reduces to the familiar ``overlap" $F(\rho,\kb\phi\phi) = \bra\phi\rho\ket\phi$. 

It is instructive to first consider the case where the measurement are perfectly accurate, $\sigma =0$ in Eq.(\ref{ex_prob}), which implies $q_\bL(\el) = \delta(\el-\bL)$ and $G(\bL,\bL') = \delta_{\bL \bL'}$.  In this case, the post measurement density matrix is
\begin{equation}
\rho'_N = \sum_{\bL} m(\bL,\bR) \ \kb{\bL}{\bL},
\label{eq:rho_post_perfect}
\end{equation}
it has completely decohered in the type basis $\ket \bL$, i.e. there are no off-diagonal terms of the form $\kb{\bL}{\bL'}$ like in Eq.~(\ref{eq:rho_post}). The fidelity between the pre- and post-measurement state is then
\begin{eqnarray}
F_{\sigma=0}(\rho_N,\rho_N') &=& \sum_\bL \left[m(\bL,\bR)\right]^2 \nonumber\\
&\leq& \underbrace{\sum_\bL m(\bL,\bR)}_1 \times \left(\max_{\bL}m(\bL,\bR)\right), \nonumber \\
&\approx& \frac{1}{(2\pi N)^{\frac{d-1}{2}} \prod_j|\beta_j|}
\end{eqnarray} 
where the subscript $\sigma=0$ indicates that the measurement are perfectly accurate, and we have used Stirling's approximation in the last line. Clearly, {\em exact type measurements greatly disturb the system}, since fidelity goes to zero as the size of the sample increases, except in the case where $\beta_j = \delta_{j_0}$.  A similar conclusion based on different considerations was reached by Squires~\cite{Squires1990}. It follows from the concavity of fidelity $F(\rho,\sum_j p_j \nu_j) \geq \sum_j p_j F(\rho,\nu_j)$ that the conditional post-measurement state $\rho_{N|\el}$ has, with high probability, a vanishing fidelity with the original state $\rho_N$.   

The disturbance caused by an exact type measurement is most obvious when considering the conditional post-measurement state of a {\em single} molecule from the sample.  As shown in the Appendix, 
\begin{equation}
\rho_{1|\el} = \sum_{j=1}^d \ell_j \kb{x_j}{x_j} \ :
\label{eq:single_post}
\end{equation} 
the conditional state of a single molecule is diagonal in the $\ket{x_j}$ basis with eigenvalues given by the {\em observed} type of the sample $\el$, independently of its state $\nu$ prior to the measurement. However, following the typical sequence theorem Eq.~(\ref{eq:tst}), the observed coefficients $\ell_j$ are very likely to be close to $\bra{x_j}\nu\ket{x_j}$. When averaging over measurement outcomes, we recover the state $\nu \rightarrow \rho_{1|\el} = \sum_j R_j \kb{x_j}{x_j}$ which has no off-diagonal terms, i.e. $\kb{x_i}{x_j}$. Thus, the exact measurement of a macroscopic observable {\em completely decoheres individual molecules of the sample}; it leaves the diagonal elements of $\nu$ unchanged while suppressing all off-diagonal terms. (This situation might appear worrisome for bulk-state quantum computing; we will return to this in Section~\ref{NMR}). Moreover, the measurement creates correlation between the molecules, so  $\rho_{N}' \neq (\rho_{1}')^{\otimes N}$ and $\rho_{N|\el} \neq (\rho_{1|\el})^{\otimes N}$ in general.

\subsection{Gaussian smoothing}

We now turn our attention to the case where the smoothing function $q_\bL(\el)$ has a finite width $\sigma$. In the case of interest, the initial state of the sample $\ket{\Psi_N}$ is pure, so combining Eqs.~(\ref{eq:state_copy}) and (\ref{eq:rho_post}) we get
 \begin{eqnarray}
F(\rho_N,\rho_N') &=& \bra{\Psi_N}\rho_N'\ket{\Psi_N} \label{eq:fidelity}\\
&=& \sum_{\bL,\bL'} m(\bL,|\beta_j|^2)m(\bL',|\beta_j|^2) G(\bL,\bL').\nonumber
\end{eqnarray} 
For sake of definiteness, we will consider the Gaussian distribution $q_\bL(\el)$ defined at Eq.(\ref{ex_prob}). The decoherence kernel defined at Eq.~(\ref{eq:kernel}) is then given by
\begin{eqnarray*}
G(\bL,\bL') &=& \int \left(\frac{1}{2\pi\sigma^2}\right)^{\frac d2}
e^{-\frac{\|\el-\bL\|_2^2+\|\el-\bL'\|_2^2}{4\sigma^2}} d\el \\
&=& \exp\left\{-\frac{\|\bL-\bL'\|_2^2}{2(2\sigma)^2}\right\} .
\end{eqnarray*}
This is not surprising as the decoherence kernel is the convolution of the smoothing function with itself. The convolution of two distribution of width $\sigma_1$ and $\sigma_2$ gives a distribution of width $\sigma = \sigma_1+\sigma_2$, so $G(\bL,\bL')$ is a function of width $2\sigma$.  

We can find a lower bound to the fidelity by truncating the sum in Eq.(\ref{eq:fidelity}). By restricting $\bL$ and $\bL'$ to the domain $\mathcal{D} = \{\bL: \|\bL -{\bf R}\|_2 \leq \Delta\}$ where $R_j = \bra{x_j}\nu\ket{x_j}$, we can lower bound the kernel by $G(\bL,\bL') \geq \exp\{-\frac{\Delta^2}{2\sigma^2}\}$ using the triangle inequality. This yields the inequality
\begin{equation*}
F(\rho_N,\rho_N') \geq \exp\left\{-\frac{\Delta^2}{2\sigma^2}\right\} 
\left(\sum_{L \in  \mathcal{D}} b(L)\right)^2.
\end{equation*}
The quantity in the parenthesis is a sum over the range $\mathcal D$ of a multinomial probability distribution. It is equal to $P(\|\bL-{\bf R}\|_2 \leq \Delta) \geq P(\|\bL-{\bf R}\|_1 \leq d \Delta) \geq (1-e^{-Nd\Delta/2})$ by the Cauchy-Schwartz inequality and the typical sequence theorem Eq.(\ref{eq:tst}). Thus, we get
\begin{equation}
F(\rho_N,\rho_N') \geq \exp\left\{-\frac{\Delta^2}{2\sigma^2}\right\} \left(1-e^{-Nd\Delta^2/2}\right)^2.
\label{eq:lowerbound}
\end{equation}
Since this bound holds for all $\Delta$ (which is an arbitrary cut-off), we can maximize the RHS of Eq.(\ref{eq:lowerbound}) --- the optimal value turn out to be attained when $\Delta^2 = 2\ln(1+2N\sigma^2d)/Nd$ --- to get the tightest bound:
\begin{equation}
F(\rho_N,\rho_N') \geq 
1-\frac{1+\ln(2N\sigma^2d)}{N\sigma^2d}.
\label{eq:bound_gauss}
\end{equation}
Hence, the measurement accuracy $\sigma$ can decrease as fast as $1/\sqrt N$ while maintaining a constant fidelity $F(\rho_N,\rho_N') = 1-\epsilon$ between the pre- and post-measurement states. If $\sigma$ decreases less rapidly than $1/\sqrt N$, e.g. $N^{-s}$ for $0<s<1/2$, the fidelity will go to $1$ as $N$ grows. In particular, if $\sigma$ is constant, $F(\rho_N,\rho_N') \sim 1-c\frac{\ln N}{N}$. 

The fidelity between the pre- and {\em conditional} post-measurement state, i.e. $\rho_N$ and $\rho_{N|\ell}$ respectively, can be computed using similar techniques. The computation is illustrated in the Appendix for two-dimensional molecules; the general case is as conceptually straightforward as it is notationally cumbersome. In this case, even when $\sigma \gg 1/\sqrt N$, not all measurement outcomes $\tilde Q_\ell^{(N)}$ yield a state $\rho_{N|\ell}$ ``close" to the initial state. However, with a probability which approaches unity as $1-\exp\{-\sqrt{\log N}\}$, fidelity will approach unity as $1-F(\rho_N,\rho_{N|\ell}) \propto \exp\{-\sqrt N\}$. In short, as $N$ increases, the measurement induces a negligible disturbance to the state of the ensemble except with a vanishing probability. 

\subsection{General smoothing}\label{general}

We now wish to argue that the essence of our measurement accuracy-state disturbance tradeoff applies to arbitrary smoothing function $q_{\bf L}({\el})$ introduced at Eq.(\ref{eq:cg_proj}), provided that it is actually smooth with respect to $\bL$. Let us be more precise. Intrinsic to the smoothing function is a notion of distance on the real $d$-dimensional vector space. One can define various distance measure on this space, e.g. our choice of smoothing function Eq.~(\ref{ex_prob}) in the previous section relied on the distance $\|\bL,\bL'\|_2^2$ induced by the $L_2$-norm. The exact statement of the tradeoff will obviously depend on the choice of distance measure. However, the essence of the result is independent of this choice, as all good distance measure are equivalent on small distances. Thus, a good smoothing function $q_\bL(\el)$ should satisfy
\begin{equation}
|q_\bL(\el)-q_{\bL'}(\el)| \leq c \left(\frac{\|\bL,\bL'\|_1}{\sigma}\right)^s
\label{eq:smooth}
\end{equation}
for sufficiently small $\|\bL-\bL'\|_1$ and some positive constants $c$ and $s$ (Eq.~({\ref{eq:smooth}) is known as Lipschitz condition). In general, $c$ depends on the dimension $d$ of the molecules. Therefore, the dependence of the bound Eq.(\ref{eq:bound_gauss}) on the dimension $d$ (which may seem counterintuitive) only reflects our choice of the $L_2$-norm in the smoothing function, it is not universal. Given this assumption, we can derive the general result. It should be mentioned that ultimately, $q_\bL(\el)$ depend on the details of the measurement procedure of the corresponding macroscopic observable (see for example the model of Section~\ref{NMR}). However, if this measurement is of finite accuracy, then the smoothing function must have a certain width and should satisfy the above assumption.

We see from Eq.(\ref{eq:fidelity}) that fidelity between the pre- and post-measurement state only depends on the decoherence kernel $G(\bL,\bL') = \int \sqrt{q_\bL(\el)q_{\bL'}(\el)}d\el$. Thus, the procedure used in the previous section carries through straightforwardly. We can truncate the sum Eq.(\ref{eq:fidelity}) to the domain $\mathcal D$ where $\|\bL -{\bf R}\|_1 \leq \Delta$. On this domain, the fluctuations of the kernel are bounded by Eq.(\ref{eq:smooth}) using the triangle inequality. Moreover, as $G(\bL,\bL) =1$ by the normalization condition of the smoothing function, we obtain
\begin{equation*}
G(\bL,\bL') \geq 1- c(\Delta/\sigma)^s \rm{\ on\ } \mathcal{D}.
\end{equation*}
The bound 
\begin{equation}
F(\rho_N,\rho_N') \geq \left\{1-c\left(\frac{\Delta}{2\sigma}\right)^s\right\} (1-e^{-N\Delta^2/2}).
\end{equation}
follows straightforwardly from the typical sequence theorem Eq.(\ref{eq:tst}). Given the value of $c$ and $s$, one can perform an optimization with respect to $\Delta$ to get the tightest bound. However, this depends on the details of the smoothing function.

Similarly, we can derive a bound for the fidelity of the conditional post-measurement state when $\sigma \gg 1/\sqrt N$. In this case, the multinomial distribution $m(\bL,\bR)$ behaves like a Kronecker delta $\delta_{\bL,{\bf R}}$ on the scale $\sigma$, so following Eq.~(\ref{eq:probability_ell}), $P(Q_\el^{(N)}|\rho_N)  \approx q_\bR(\el)$. The fidelity between the pre- and conditional post-measurement state is therefore
\begin{equation*}
F(\rho_N,\rho_{N|\el}) \approx \sum_{\bL,\bL'} m(\bL,|\beta_j|^2)m(\bL',|\beta_j|^2) \frac{\sqrt{q_\bL(\el)q_{\bL'}(\el)}}{q_\bR(\el)}.
\end{equation*}
Again, we can truncate the sum to the domain where $\|\bL -{\bf R}\|_1 \leq \Delta$ to get the bound
\begin{equation*}
F(\rho_N,\rho_{N|\el}) \geq \left(1-\frac{c(\Delta/\sigma)^s}{q_\bR(\el)}\right)(1-e^{-N\Delta^2/2})
\end{equation*}
which goes to one with probability one as $N$ increases, as illustrated in the Appendix for a specific choice of smoothing function. 

Finally, the scaling $\sigma \sim 1/\sqrt N$ is optimal. A higher precision would considerably disturb the state of the system. This is because the multinomial distribution $m(\bL,\bR)$ has a width $1/\sqrt N$. Consider the expression of  Eq.(\ref{eq:fidelity}). If the kernel has a width smaller than the binomial distribution, the sum, and hence the fidelity $F(\rho_N,\rho_N')$, will be roughly equal to erf$(\sigma\sqrt N) \approx 2\sigma/\sqrt{N\pi}$ for $\sigma \ll 1/\sqrt N$. 
The bound is also tight for the conditional post-measurement fidelity $F(\rho_N,\rho_{N|\el})$ as fidelity is a convex function. This can also be seen intuitively by considering the behavior of two consecutive measurements. Upon fine grained measurement $Q^{(N)}_\bL$, the variance of the outcome $\bL$ is $1/\sqrt N$. However, if we first perform a coarse grained measurement $\tilde Q^{(N)}_\el$ of width $\sigma \ll 1/\sqrt N$ and then perform a fine grained measurement on the updated state $\rho_{N|\ell}$, the variance of the second measurement outcome will be $\sigma$: performing the coarse grained measurement has altered its statistics. This means that the coarse grained measurement has appreciably disturbed the state of the sample, so $F(\rho_N,\rho_{N|\el})$ is far from 1. 

\subsection{Mixed states}

The result of the previous section hold unchanged when the molecules of the sample are all prepared in the same {\em mixed} state $\nu = \sum_{i=1}^d \lambda_i \kb{\psi_i}{\psi_i}$. The argument proceeds in three steps. First, we can construct a purification of the state $\nu$ 
\begin{equation*}
\ket{\phi} = \sum_{i=1}^d \sqrt{\lambda_i} \ket{\psi_i}\ket i
\end{equation*}
by appending to each molecule an ancillary system of dimension $d$ with orthonormal basis $\{\ket i\}$. Clearly, the reduced state of the molecule --- obtained by tracing out the ancilla --- is $Tr_{\mathrm{ancilla}}\{\kb\phi\phi\} = \nu$. Second, the vectors $\{\ket{x_j}\ket i\}_{i,j = 1,\ldots,d}$ form a basis for the Hilbert space of the pair molecule+ancilla. The type projectors $Q_\bL^{(N)}$ associated to the molecule only measure the type of the prefix $x_j$, so are coarse grained version of the type projectors associated to the pair: the disturbance they cause to the state of the sample can only be less than the disturbance caused by the complete type projectors. Thus, the bound Eq.(\ref{eq:lowerbound}) can be applied to $F(\Phi_N,\Phi_N')$ where $\Phi_N = \kb{\phi}{\phi}^{\otimes N}$, and 
\begin{equation*}
\Phi_N' = \int (\tilde Q_\el^{(N)}\otimes\II) \Phi_N (\tilde Q_\el^{(N)}\otimes\II) d\el.
\end{equation*}
Finally, by monotonicity of the fidelity --- $F(\mathcal{E}(\rho),\mathcal{E}(\nu)) \geq F(\rho,\nu)$ for any trace preserving quantum operation $\mathcal E$ --- the bound applies directly to the pre- and post-measurement state of the sample of molecules by tracing out the ancillas. By similar considerations, all of the above conclusions can be extended to mixed states.

\section{Exchangeability}\label{exchangeability}

Before proceeding with the applications of the above results, we wish to introduce a wider class of states --- exchangeable states --- to which our results can be applied. The concept of {\em exchangeability} was introduced in the classical theory of probability by de Finetti~\cite{deFinetti} to substitute the incorrect use of ``unknown probabilities''. A probability assignment is the expression of one's subjective knowledge about the possible outcomes of an experiment. Hence, it is not a property of a physical system itself but, rather, a property of the agent assigning the probability, so it can not be unknown to him!

There are also several good reasons to believe that quantum states are subjective, see for example \cite{CF1996,FP2000,Mermin2001,Fuchs2002} and references therein. The state of a quantum system is a mathematical construct which allows one to compute probabilities for various measurements outcomes.\footnote{To quote Robert Griffiths, ``{\em If probabilities are not real, then pre-probabilities} [quantum states] {\em are even less real}" \cite{Griffiths}.} As a consequence of the subjective nature of quantum states, the concept of an {\em unknown quantum state} is in general an oxymoron, for essentially the same reasons which lead to this conclusion for classical probability assignment. 

However, unknown quantum state turn out to be quite useful for the description of certain physical setting. Of particular interest to us is the description of a sample of $N$ ``molecules". Under certain circumstances --- e.g thermal equilibrium --- one can arrive at the conclusion that all the molecules of the sample are equivalent, so they should all be describe by the same state $\nu$, which is itself unknown. This is a very common state of affair in nuclear, atomic, or molecular physics where spectral quantities --- which are formally described by macroscopic observables --- are measured over a large collection of quantum systems. In fact, in almost all physical experiment where ensemble measurements are performed, the components of the sample are assumed to be in the ``same unknown state", and the purpose of the measurement is to (partially) determine this state. 

To arrive at an appropriate description of the sample without refering to the unknown quantum state of individual molecules, we must clearly state the assumption of the agent assigning the state. His assumption is that {\em the arbitrary number of molecules are all equivalent}, which can be formalized as follows
\begin{enumerate}
\item For any permutation $\pi$ of the $N$ molecules, $\pi[\rho] = \rho$. Such a state is called symmetric.
\item For any positive integer $M$, there exists a symmetric state $\rho_{N+M}$ such that $\rho_N = Tr_M \{\rho_{N+M}\}$, where $Tr_M$ denotes the partial trace over $M$ molecules. 
\end{enumerate} 
A state $\rho$ satisfying these two conditions is called {\em exchangeable}. The quantum de Finetti representation theorem \cite{HM1976,Hudson1981,CFS2002} asserts that any exchangeable quantum state $\rho_N$ of a sample of $N$ molecules can be written as 
\begin{equation}
\rho_N = \int \nu^{\otimes N} Pr(\nu) d\nu
\label{eq:exchangeable}
\end{equation}
where $\nu$ are density operators of a single molecule and $Pr(\nu)$ is a probability distribution over the quantum states of a single molecule. 

The interpretation of this theorem is that it is {\em mathematically correct} to look upon $\nu$ as an objective element of reality about which we have incomplete knowledge: hence we assign it some probability distribution $Pr(\nu)$.  For example, when the POVM $\{E_i\}$ is measured on the sample, the outcome $E_j$ is observed with probability
\begin{eqnarray}
P(E_j|\rho_N) &=& Tr\{E_j \rho_N\} \nonumber \\
&=& \int Tr\{E_j \nu^{\otimes N}\} Pr(\nu) d\nu \nonumber \\
&=& \int P(E_j|\nu^{\otimes N})Pr(\nu) d\nu.
\label{eq:prob_exchangeable}
\end{eqnarray}
We can think of $P(E_j|\nu^{\otimes N})$ as the probability of $E_j$  given a value of the {\em real parameter} $\nu$, but since $\nu$ is unknown, we average this probability over the possible values of $\nu$ distributed according to $Pr(\nu)$. However, it must be emphasized that it is the assumption of exchangeability which leads to the form of Eq.(\ref{eq:exchangeable}), which in turn legitimizes the term ``unknown state" for mathematical convenience.

\subsection{Bulk tomography}\label{tomography}

Quantum state tomography is an experimental procedures which transforms an exchangeable state of the form Eq.(\ref{eq:exchangeable}) to a product state $\rho = \nu^{\otimes N}$. According to the de Finetti representation theorem, we can equivalently say --- and this is how tomography is conventionally formulated --- that the purpose of tomography is to determine which is the {\em real yet unknown} state $\nu$ describing the $N$ molecules of the sample. 

In \cite{CFS2002}, quantum state tomography was studied in the context where the molecules of the sample are measured individually. Here, we present how quantum state tomography can be performed through bulk measurements. A similar description was recently and independently developed in~\cite{BHL2004}. Let $A_N = \sum_k a_{(k)}$ be a macroscopic observable deriving from the single-molecule observable $a$ as in Eq.(\ref{eq:macroscopic}). Exceptionally, we denote the eigenstates and eigenvalues of $a$ with a superscript $a\ket{x_j^{(a)}} = \alpha_j^{(a)}\ket{x_j^{(a)}}$ for later convenience. The finite accuracy measurements of the macroscopic observable $A_N$ is defined through the POVM $\{\tilde Q^{(N,a)}_\el\}$. 

The conditional state of the sample after the measurement of $\{\tilde Q^{(N,a)}_\el\}$ with outcome $\el^{(a)}$ is 
\begin{equation*}
\rho_{N|\el^{(a)}} = \int \left(\sqrt{\tilde Q^{(N,a)}_\el} \nu^{\otimes N}  \sqrt{\tilde Q^{(N,a)}_\el}\right) \frac{Pr(\nu)}{P(\tilde Q^{(N,a)}_\el|\rho_N)}d\nu .
\end{equation*}
The quantity in parenthesis is proportional to the conditional post-measurement state of the sample, given that it was initially in state $\nu^{\otimes N}$. As demonstrated in Section~\ref{main}, this measurement has, except with a vanishing probability, very high fidelity with the original state. Mathematically, this means
\begin{equation*}
\sqrt{\tilde Q^{(N,a)}_\el} \nu^{\otimes N}  \sqrt{\tilde Q^{(N,a)}_\el} \approx P(\tilde Q^{(N,a)}_\el |\nu^{\otimes N}) \nu^{\otimes N}
\end{equation*}
with probability approaching unity as $N$ grows. Therefore, we get
\begin{equation*}
\rho_{N|\el^{(a)}} \approx \int \nu^{\otimes N}  \frac{Pr(\nu)P(\tilde Q^{(N,a)}_\el |\nu^{\otimes N})}{P(\tilde Q^{(N,a)}_\el|\rho_N)}d\nu .
\end{equation*}
This is {\em as if} we had updated the probability distribution $Pr(\nu)$ of the real yet unknown state $\nu$ according to Bayes' rule $P(y|x) = P(x|y)P(y)/P(x)$. However, this is strictly a mathematical identity, all we did was to apply the state update rule Eq.(\ref{eq:general_update}) to an exchangeable state. 

We can repeat the procedure with a different macroscopic observable $B_N, C_N, \ldots$ derived from the single-molecule observable $b,c,\ldots$, which do not necessarily commute with each other. If the sets of observables are sufficiently informative --- i.e. if $\{\kb{x_j^{(k)}}{x_j^{(\mu)}}\}_{\mu=a,b,c,\ldots}$ contain $d^2-1$ linearly independent elements --- the updated probability distribution will converge with very high probability to a delta function for sufficiently large $N$, $Pr(\nu |\el^{(a)},\el^{(b)},\ldots) \approx \delta(\hat\nu)$, so the final state will be $\rho_{N|\el^{(a)}\el^{(b)}\ldots} \approx \hat\nu^{\otimes N}$. This is because the functions $P( \hat Q^{(N,\mu)}_\el | \nu^{\otimes N)}$ are centered around $\el^{(\mu)} = (\bra{x_1^{(\mu)}}\nu\ket{x_1^{(\mu)}}, \bra{x_2^{(\mu)}}\nu\ket{x_2^{(\mu)}}, \ldots)$ and have a width $\sigma$. The state $\hat\nu$ is the only one satisfying all the linear constraints $\bra{x_j^{(\mu)}}\hat\nu\ket{x_j^{(\mu)}} = \ell_j^{(\mu)}$ for all $\mu=a,b,c,\ldots$ up to accuracy $\sigma$.  Again, this is ``as if" the measurements simply inform us of the identity of the ``real" $\hat\nu^{\otimes N}$, without disturbing it in the limit of large $N$. 

\section{Classicality}\label{classicality}

The purpose of this section is to demonstrate that macroscopic observables define a classical limit for closed quantum systems, i.e.  it does not require interaction with any ``environment". Observations play very different roles in classical and quantum theory. In the classical setting, we can think of measurements as unveiling an underlying ``real" state of affairs: observations reveal information about the state of the world without affecting it. On the other hand, quantum measurements disturb or ``collapse" the state of the system.

When states, either quantum or classical, are regarded as subjective judgments of the world, both of the above descriptions need revision. Let $P(x_i,y_j)$ be the joint probability distribution, or state, that the agent assigns to the classical sequence of events $X=x_i$ and $Y=y_j$. Upon the observation $X=x_i$, the agent updates her predictions for event $Y$ according to Bayes' rule
\begin{equation*}
P(y_j|x_i) = \frac{P(x_i,y_j)}{P(x_i)}
\end{equation*}
where $P(x_i) = \sum_j P(x_i,y_j)$. This state generally differs from the pre-measurement state assigned to $Y$
\begin{equation*}
P(y_j|x_i) \neq P(y_j) = \sum_i P(x_i,y_j).
\end{equation*}
Hence, the act of observing $X$ modifies the state assigned to $Y$.  However disregarding the observed value of $X$  for later probability assignments is like not measuring the value of $X$ at all:
\begin{equation}
P(y_j) = \sum_i P(y_j|x_i)P(x_i).
\label{eq:bayes}
\end{equation}
Indeed, we can interpret the observation as revealing the ``real" value of $X$ which was there all along: the agent simply didn't know about it prior to her observation. In this sense, $X=x_i$ is a real state of affair about which the agent learns through the act of measurement. Thus, the state she assign to $Y$ prior to her observation of $X$ is the mixture of the state it would have given the different value of $X$, weighted by the probability of $X$, c.f. Eq.~(\ref{eq:bayes}). This reasoning extends in an obvious way to any sequence of events $X^{(1)},X^{(2)},\ldots,X^{(n)}$. We can consider that the system follows a fixed history $x_{j_1}^{(1)},x_{j_2}^{(2)},\ldots,x_{j_n}^{(n)}$ of which the agent has incomplete knowledge.

Quantum measurements behave quite differently. A quantum event corresponds to a ``click" on a measurement apparatus at some instant of time $t$. Hence, each event is associated a POVM element $E_{j_k}^{(k)}(t_k)$ in the Heisenberg picture at a given time $t_k$. (We will henceforth drop the explicit time label $t_k$.) In general, assigning definite yet unknown outcomes to these events lead to incorrect predictions, e.g. the sum rule
\begin{equation*}
P(E_{j_2}^{(2)}) = \sum_{j_1} P(E_{j_2}^{(2)}|E_{j_1}^{(1)}) P(E_{j_1}^{(1)})
\end{equation*}
does not hold in general. This is most obvious in Young's double slits experiment where the events $E^{(1)}_{j_1}$ correspond to the particle going through slit $j_1 =$ 1 or 2 and $j_2$ labels the various positions on the detector. Reasoning involving the particle going through a definite yet unknown slit lead to incorrect predictions. 

There are however sequences of quantum events which behave classically, as if the observations were revealing an underlying reality. The typical example being when all the POVM elements describing the events commute. The consistent histories approach to quantum theory~\cite{histories} lays down a set of conditions under which such behavior occurs. A complete list of alternatives events $\zeta^{(k)} = \{E_{j_k}^{(k)}\}$ at time $t_k$ defines a POVM. A history is a list of POVM elements $H = (E_{j_1}^{(1)},E_{j_2}^{(2)},\ldots,E_{j_n}^{(n)})$ at distinct times $t_1,t_2, \ldots ,t_n$. When the initial state of the system is $\rho$, the probability of an history $H$ is 
\begin{equation}
P(H|\rho) = Tr\left\{(E_{j_n}^{(n)})^{\frac 12}\ldots (E_{j_1}^{(1)})^{\frac 12} \rho (E_{j_1}^{(1)})^{\frac 12} \ldots (E_{j_n}^{(n)})^{\frac 12} \right\}
\label{eq:probH}
\end{equation}
following Eqs.(\ref{eq:general_prob},\ref{eq:general_proj}). 
A complete family of histories is the set of all combination of POVM elements from the sets $\zeta^{(k)}$ at all times, $\mathcal F = \{\zeta^{(1)},t_1;\zeta^{(2)},t_2;\ldots \zeta^{(n)},t_n\}$. A family is thus a sample space on which a probability distribution $P(H|\rho)$ is defined. The familly is said to be {\em consistent} when the sum rule approximately holds for $P(H|\rho)$. This conditions is the simplest version of all consistency conditions but will be sufficient for our purposes. In this sense, consistent histories define a quasiclassical domain of familiar experience. 

As was observed by Halliwell~\cite{Halliwell}, histories corresponding to a sequence of finite accuracy measurement of macroscopic observables generate a consistent family if the system is a sufficiently large sample of identically prepared molecules, i.e. when $\rho = \nu^{\otimes N}$. In other words, one can measure any sequence of macroscopic observable on a large sample and account for the observed statistics with a classical theory. It should be stressed that the single-molecule observables $a,b,c,\ldots$ making up the histories {\em do not need to commute}. For example, the coarse measurement of the magnetization of a sample of spin-$\frac 12$ along the $z$ axis followed by a measurement along the $y$ axis can generate a consistent family if the sample is sufficiently large.

An alternative way of building our intuition in this direction is to consider the commutator of any two {\em normalized} macroscopic observables. Let $a$ and $b$ be two arbitrary single-molecule observables and define $c$ to be their commutator $c=[a,b]$. These operator can be suitably normalized so that they satisfy $\|a\|,\|b\|,\|c\| \approx 1$. The normalized macroscopic observable $A_N$ is defined as $A_N = \frac 1N \sum_{k=1}^N a_{(k)}$, and similarly for $B_N$ and $C_N$; hence $\|A_N\|, \|B_N\|,\|C_N\| \approx 1$. A straightforward calculation shows that the commutator of the normalized macroscopic observables obeys
\begin{equation}
[A_N,B_N] = \frac 1N C_N, 
\label{eq:commutation}
\end{equation} 
which implies $\|[A_N,B_N]\| \approx \frac 1N$. Thus, all macroscopic observables commute in the limit of infinite-size sample, and commuting observables systematically generate consistent histories: measuring the value of one observable does not affect the outcome statistics of other commuting observables. Once again, the infinite-sample considerations can not be applied straightforwardly to finite ensembles (this is the recurrent theme of this paper!) In particular, Eq.~(\ref{eq:commutation}) does not involve any coarseness, which is essential to achieve consistency in finite ensembles. 

The conclusion reached by Halliwell can be extended to a much wider class of initial states. The first generalization is straightforward: by linearity of Eq.(\ref{eq:probH}), such families are automatically consistent for initial exchangeable states. Indeed, for initial state $\rho_N$ of the form Eq.(\ref{eq:exchangeable}), the probability of history $H$ reads
\begin{equation}
P(H|\rho_N) = \int P(H|\nu^{\otimes N}) Pr(\nu) d\nu.
\end{equation}
Clearly, if the sum rule is satisfied for the $P(H|\nu^{\otimes N})$ individually, it is also satisfied for their convex combination. This is very much in the spirit of the de Finetti representation theorem as one can interpret the outcome of the macroscopic measurements as revealing partial information about the {\em real} quantum state $\nu^{\otimes N}$ of the sample, of which we have incomplete knowledge.  

Moreover, consider an arbitrary product state of the sample $\rho = \nu_1\otimes\nu_2\otimes\ldots\otimes\nu_N$.  We will construct a state $\overline\nu^{\otimes N}$ whose measurement outcomes, for coarse grained macroscopic observables, are statistically indistinguishable from those obtained from the product state $\rho$. This will prove that Haliwell's result applies to arbitrary product states. Consider the symmetrized version of $\rho$: 
\begin{equation*}
\Pi[\rho] = \frac{1}{N!} \sum_\pi \nu_{\pi(1)}\otimes\nu_{\pi(2)}\otimes\ldots\nu_{\pi(N)}
\end{equation*}
where the sum is over all permutations of $N$ elements. The reduced state of a single molecule is 
\begin{equation}
Tr_{N-1}\{\Pi[\rho]\} = \frac 1N \sum_{k=1}^N \nu_k = \overline\nu. 
\end{equation}
The states $\Pi[\rho]$ and $\overline\nu^{\otimes N}$ are in some sense very similar: they are both symmetric, they yield the same reduced single-molecule state $\overline\nu$, and yield the same expectation value of the frequency operator $\langle {\bf F}^{(N)}\rangle = (\bra{x_1}\overline\nu\ket{x_1},\bra{x_1}\overline\nu\ket{x_1},\ldots)$. However, they are not identical. To illustrate this, consider a sample of $N$ two-dimensional molecule in the state
\begin{equation*}
\rho = \underbrace{\kb{x_1}{x_1}\otimes\ldots\otimes\kb{x_1}{x_1}}_{N/2}\otimes
\underbrace{\kb{x_2}{x_2}\otimes\ldots\otimes\kb{x_2}{x_2}}_{N/2}:
\end{equation*}
half of the molecules are in the state $\ket{x_1}$ while the other half are in state $\ket{x_2}$. 
The measurement of the frequency operator ${\bf F}^{(N)}$ of Eq.(\ref{eq:frequency}) yields outcome $(\frac 12,\frac 12)$ with certainty when the state of the system is $\Pi[\rho]$.  The average result will also be $(\frac 12,\frac 12)$ when the sample is in state $\overline\nu^{\otimes N}$, but can fluctuate around this value. However, according to the typical sequence theorem Eq.(\ref{eq:tst}), the size of these fluctuations will be of order $1/\sqrt N$, so can only be perceived by macroscopic measurements of accuracy $\sigma \lesssim 1/\sqrt N$. For macroscopic observables of coarseness $\sigma \gg 1/\sqrt N$, the two states $\Pi[\rho]$ and $\overline\nu^{\otimes N}$ will yield the same statistics up to order $\sigma\sqrt N \ll 1$.  

Moreover, the states $\rho$ and $\Pi[\rho]$ yield exactly the same statistics for measurement outcome of macroscopic observables: this follows straightforwardly from the permutation invariance of the type projectors Eq.(\ref{eq:proj_pop}). We have thus established the chain of equality
\begin{equation*}
P(\tilde Q_\el^{(N)}|\rho) = P(\tilde Q_\el^{(N)}|\Pi[\rho]) \approx P(\tilde Q_\el^{(N)}|\overline\nu^{\otimes N}),
\end{equation*}
so the states $\rho$ and $\overline\nu^{\otimes N}$ yield almost identical predictions when $\sigma \gg 1/\sqrt N$. It follows that a sequence of finite accuracy macroscopic measurements performed on a state of the form $\rho = \nu_1\otimes\nu_2\otimes\ldots\otimes\nu_N$ generate a consistent family of histories for sufficiently large samples. 

In fact, {\em any separable state leads to consistency of macroscopic histories}. Indeed, when the molecules of the sample are {\em not entangled} with each other, their state can be written as
\begin{equation*}
\rho = \int \nu_1\otimes\nu_2\otimes\ldots\otimes\nu_N Pr(\nu_1,\nu_2,\ldots,\nu_N) d\nu_1 d\nu_2\ldots d\nu_N.
\end{equation*}
Now, consider the state
\begin{equation}
\overline\rho = \int \overline \nu^{\otimes N} Pr(\overline \nu ) d\overline\nu
\end{equation}
where we have defined
\begin{equation*}
Pr(\overline\nu) = \int Pr(\nu_1,\ldots,\nu_N) \delta\Big(\overline\nu-\frac 1N \sum_k \nu_k\Big) d\nu_1\ldots d\nu_N. 
\end{equation*} 
By linearity and using the result established above, we have $P(H|\rho) \approx P(H|\overline\rho)$ provided that the histories $H$ are generated by macroscopic observable of accuracy $\sigma \gg 1/\sqrt N$. Since $\overline\rho$ is exchangeable, the probabilities $P(H|\overline\rho)$, and therefore the probabilities $P(H|\rho)$, approximately satisfy the sum rule. 

To put it simply, when macroscopic measurements are coarse with respect to the quantum correlation length scale of the system, they behave classically. Indeed, assume that a sample of $N$ molecules have correlation length $\xi$, i.e. there is $\xi$-molecule entanglement in the system. Then, all of the above construction can be applied to the $N/\xi$ collections of $\xi$ molecules. We simply have to treat each block of entangled $\xi$ molecules as one big molecule.  There is no entanglement between these large molecules so the previous analysis applies, as long as the measurement accuracy is larger than $\sqrt{\xi/N}$. Thus, we see that only entanglement on ``macroscopic" scales can cause quantum effects to the measurement of coarse grained macroscopic observable.   

\section{NMR information processing}\label{NMR}

Room temperature nuclear magnetic resonance (NMR) has been for several years a benchmark for quantum information processing~\cite{NMR}. The sample contains $N\approx10^{20}$ molecules which are to good approximation non interacting due to dynamical decoupling caused by thermal excitations. Hence, the total Hamiltonian is the sum of the single-molecules hamiltonian $H = \sum_k h_{(k)}$, it takes the form Eq.(\ref{eq:macroscopic}), so it is a macroscopic observable. Initially, the sample is in a thermal state 
\begin{equation*}
\rho_N = \frac{e^{-\beta H}}{Z} = \frac{e^{-\sum_k h_{(k)}}}{Z} = \left(\frac{e^{-\beta h}}{z}\right)^{\otimes N}
\end{equation*}
where $Z = Tr\{e^{-\beta H}\}$ and $z = Tr\{e^{-\beta h}\}$ are the partition functions of the sample and of a single molecule respectively. Each molecule contains a certain number of nuclei which carrie a spin, and it is these spin degree of freedom which are used to perform the computation. The various spins of a molecule can have different Larmor frequencies $\omega_j$, which makes it possible to address them individually. To do so, the sample is placed by a coiled wire through which a sequences of externally controlled radio frequency (RF) current pulses can be applied. By properly tuning the frequency of the RF pulse, we can address all the spins with the same Larmor frequency, so all the $N$ molecules are addressed in parallel. Therefore, a sequence of pulses transforms the state of the sample according to
\begin{equation*}
\rho_N \rightarrow U^{\otimes N} \rho_N U^{\dagger \otimes N},
\end{equation*}
which preserves the tensor product structure of the density matrix $\rho_N = \nu^{\otimes N}$, it collectively changes the state of individual molecules $\nu$.\footnote{When the sequence of pulses generates a complex transformation, it is practically impossible to keep tract of $\nu$, as this would require an exponential amount of computation. Hence, given our limited computational capacities, the sample should really be described by an exchangeable state of the form Eq.(\ref{eq:exchangeable}). Indeed, if we can assign the sample a state of the form $\nu^{\otimes N}$ after the pulse sequences, it means that the quantum computation was useless since we are able to predict its outcome!}

It has been known for a long time \cite{BP1954} that the coupling between the nuclear spins and the coil can considerably disturb the state of the sample in certain regimes through {\em back-action}. This noise is not fundamentally irreversible, it is only due to our neglecting of high order terms in the sample's Hamiltonian. However, since the coil is also used to {\em read out} the state of the sample, it must unavoidably induce extra irreversible noise, of the kind discussed in Section~\ref{main}. This result is puzzling because the coil is present throughout the computation, not only during the measurement phase, so should in principle disturb the computation.

A simple model to study the effect of this noise was presented in Ref.~\cite{LS2000}. The current in the coil can be modeled by a continuous quantum variable $R_j = \int r_j\kb{r_j}{r_j} dr_j$ where $j$ labels the modes of the field in the coil. Each field mode $R_j$ couples to the resonant  magnetization of the sample --- i.e. to the spins of Larmor frequency $\omega_j$ --- through its conjugate momentum $P_j$, $[P_j,R_j] = i$ ($P$ is the ``generator of translations" for $R$). The coupling Hamiltonian takes the form $H_c = \gamma \sum_j P_jM_j^x$ where $\gamma$ is some coupling constant and 
\begin{equation}
M_j^x = \sum_{k=1}^N \sigma_{j(k)}^x = N\frac 12\sum_L (2L-1) Q_L^{(N)} 
\end{equation}
is the total transverse magnetization of the nuclei of Larmor frequency $\omega_j$. Assume for simplicity that each molecule contain a single spin-$\frac 12$ nucleus which couples to the field mode $R_0$. This field mode is initially in state $\ket{\phi} = \int \phi(r) \ket{r} dr$ and the sample is in state $\rho_N$. After a time $t$, the joint state of the field and the sample is
\begin{eqnarray*}
\rho(t) &=& \sum_{L,L'}\int drdr' \phi(r)\phi^*(r') Q_L^{(N)}\rho_NQ_{L'}^{(N)} \\
&&\otimes \kb{r+f(L)}{r'+f(L')}
\end{eqnarray*}
where $f(L) = \gamma t N \frac 12 (2L-1)$. If the field mode $R_0$ is subsequently observed to be in state $\ket r$, the state of the sample is updated to 
\begin{equation*}
\rho_N \xrightarrow{r} \frac{\tilde Q_{r}^{(N)} \rho_N \tilde Q_r^{(N)\dagger}}{P(r|\rho_N)}
\end{equation*}
where $\tilde Q_r^{(N)} = \sum_{L'} \phi(r-f(L'))Q_{L'}^{(N)}$ are coarse grained type measurements like those of Eq.(\ref{eq:cg_proj}). The width $\sigma$ of the smoothing function $\phi(r-f(L'))$ is the initial spread of the field mode $R_0$, divided by the coupling strength and the interaction time, $\sigma \approx \sqrt{\bra\phi R_0^2 \ket\phi - \bra\phi R_0 \ket\phi ^2}/N\gamma t $. Following the results established in Section~\ref{general}, a width $\sigma \gg 1/\sqrt{N} \approx 10^{-10}$ insures us that the measurement does not perturb the computation. 

Of course, the measurements achieved in the laboratory are much coarser than $10^{-10}$. Given the results presented in this paper, we could follow \cite{LS2000} and conclude that the presence of the coil (or the NMR measurements in general) induces a negligible disturbance to the state of the sample. However, our analysis does not apply since NMR measurements are not {\em ideal} (see Section~\ref{measurements}). This is because the coil is not in a pure state at room temperature. Consider, for example, the initial state of the coil in a Gaussian {\em mixture} of Gaussian-like field modes $\rho_F \propto \int e^{-\frac{q^2}{2\sigma^2}} \kb{\phi_q}{\phi_q} dq$ where $\ket{\phi_q} \propto \int  e^{-\frac{(r-q)^2}{4\lambda^2}} \ket r dr$ (the $\phi_q$ are like coherent states). After a coupling time $t$, the observation of the field mode in state $\ket r$ updates the state of the sample to 
\begin{equation*}
\rho_N \xrightarrow{r} \rho_{N|r} =  \int e^{-\frac{q^2}{2\sigma^2}} \tilde Q_{r-q}^{(N)} \rho_N \tilde Q_{r-q}^{(N)} dq
\end{equation*}
where $\tilde Q_r^{(N)} \propto \sum_{L'} e^{-\frac{(r-f(L'))^2}{4\lambda^2}}Q_{L'}^{(N)}$. This is the continuous version of the general state update rule Eq.(\ref{eq:general_update}) for non-ideal measurements. The corresponding POVM elements $E_r = \int e^{-\frac{q^2}{2\sigma^2}} \left[\tilde Q_{r-q}^{(N)}\right]^2$ have width $\lambda+\sigma$, which determines the accuracy of the measurement outcomes following Eq.(\ref{eq:general_prob}). However, the Kraus operators $A_{qr} = \tilde Q_{r-q}^{(N)}$ have width $\lambda$. Following Eq.(\ref{eq:general_update}), it is this width which governs the disturbance caused to the state. Thus, it is not the measurement coarseness $\lambda+\sigma$ which ultimately determines the disturbance caused to the state, but the details of the measurement process. 

It is therefore necessary to have a detailed model of the interaction between the coil and the sample to evaluate its contribution to decoherence of the state of the molecules. We suspect that, in actual NMR settings, the measurement coarseness is largely due to statistical (thermal)  fluctuations of the type of $\sigma$. However, we also suspect the coherent spread of the coil's wave function $\lambda$ to be much larger than $10^{-10}$, since coherent manipulation of the molecules appears to be possible despite the coupling to the coil. These questions, however, deserve a separate study. 

\section{Conclusion}\label{conclusion}

We have demonstrated a tradeoff between measurement accuracy and state disturbance for sample of identically prepared quantum systems. A measurement coarseness smaller than $1/\sqrt N$ causes a disturbance to the state of the system which increases as the size of the ensemble grows, which is in apparent contradiction with the infinite-copy result. However, a measurement coarseness $\sigma \gg 1/\sqrt N$ induces a negligible disturbance to the state of the sample.   

Using these results, we have argued that any sequence of macroscopic observations behave essentially classically provided that there is no large-scale entanglement in the sample. More precisely, the measurement of macroscopic observables generate consistent families of histories provided that their coarseness is larger than $\sqrt{\xi/N}$ where $\xi$ is the quantum correlation length scale of the system. We do not know however whether this scaling is optimal and this question deserves a separate study. Eq.~(\ref{eq:commutation}) indeed suggests that consistency does not require any coarse-graining as $N$ increases. The exact measurement of macroscopic observables indeed disturb the state of the sample, but might not affect the statistics of other macroscopic quantities. To illustrate this point, consider an ensemble of $N$ two-level molecules, all prepared in the state $\ket{x_1}$. Clearly, if a single molecule of the ensemble gets {\em flipped} to $\ket{x_2}$, the fidelity between the original state and this ``noisy" state is zero. In this sense, a single bit flip greatly disturbs the state of the sample. However, this single flip would have very weak repercussions on the statistics of the macroscopic observations. In fact, only a measurement of accuracy $1/N$ could detect this disturbance. The histories generated by a sequence of observations which do not disturb the system are guaranteed to generate a consistent family. However, this is not a necessary condition since some type of disturbance --- as illustrated above --- do not have perceivable effects on the statistics of macroscopic observables.

An interesting question arises from the study of the relation between exchangeable states and macroscopic observations. We have seen that applying a random permutation to the molecules in a separable state yields a state which is not exchangeable, but possesses similar characteristics. We do not know what type of operation can transform a generic quantum state into an exchangeable one. We suspect that performing a tomographically complete set of macroscopic measurements on subsets of the sample followed by a random permutation of the molecules would do the trick. Physically, this would mean that a collective coupling to the environment and a diffusion process would map any state to an exchangeable state. This would be very interesting as it would extend the reach of our classicality analysis. Moreover, understanding under what circumstances can a sample of physical systems be treated {\em as if} they were all in the same unknown state is important since this is assumed in most quantum experiments performed on macroscopic samples. 

Finally, we have related our study to a NMR measurement model introduced in \cite{LS2000}. We have extended their analysis to the case where the coil is not in a pure mode state but rather in a statistical mixture of such states, like a thermal state. In this case, there are two parameters describing the macroscopic measurements: the width $\sigma$ of the POVM elements describes the accuracy of the measurements and width $\lambda$ of the Kraus operators governs the disturbance caused to the state of the sample. Therefore, a measurement accuracy $\sigma \gg 1/\sqrt N$ does not guaranty a negligible disturbance except when the measurement is ideal. The NMR measurement process therefore deserves a detailed study. 

We thanks Charles Bennett, Carl Caves, Raymond Laflamme, Camille Negrevergne, and Harold Ollivier for stimulating discussions. This work was supported in part by Canada's NSERC.   

\section{Appendix}\label{appendix}

\subsection{Single molecule post-measurement state}

We will show how to compute the post-measurement state of a single molecule, namely Eq.~(\ref{eq:single_post}). For this, it will be useful to alter our notation a bit. In this subsection only, we will consider {\em non-normalized types}: If $\bL(X)$ denotes the normalized type of $X$, then its non-normalized type is $N\bL(X)$. Thus, for this section only, $\bL(X)$ is a $d$-component vector whose $j$th component $L_j$ equal the number of occurrences of the letter $x_j$ in $X$. Adding to the notation, for the type $\bL = (L_1,\ldots,L_d)$ of a $N$-letter string $X$, we denote by $\bL^{-x_j} = (L_1,\ldots,L_j-1,\ldots L_d)$ the type of the string obtained by removing one occurence of $x_j$ from $X$. Of course, this is a well defined type only when $L_j \geq 1$. 

Given this notation, we can write
\begin{equation}
Q^{(N)}_\bL = \sum_j \kb{x_j}{x_j}\otimes Q^{(N-1)}_{\bL^{-x_j}}
\end{equation}
where the $Q^{(N-1)}_{\bL^{-x_j}} =0$ when $\bL^{-x_j}$ is not a well defined type. Applying the state update rule and tracing out all but a single molecule, we get
\begin{eqnarray}
\rho_{1|\ell} &=& Tr_{N-1}\left\{
\frac{Q_\bL^{(N)}\rho_NQ_\bL^{(N)}}{P(Q_\bL^{(N)}|\rho_N)}\right\} \nonumber \\
&=& \sum_{ij} \kb{x_i}{x_i} \nu\kb{x_j}{x_j} 
Tr\left\{\frac{Q_{\bL^{-x_i}}^{(N-1)}\nu^{\otimes N-1}Q_{\bL^{-x_j}}^{(N-1)}} 
{P(Q_\bL^{(N)}|\rho_N)}\right\} \nonumber \\
&=& \sum_{j} R_j \kb{x_j}{x_j} \frac{P(Q_{\bL^{-x_j}}^{(N-1)}|\rho_{N-1})}
{P(Q_\bL^{(N)}|\rho_N)};
\label{eq:prob_ratio}
\end{eqnarray}
in the last line, we used the definition of the probability Eq.~(\ref{eq:prob_type}) and the orthogonality of the type projectors Eq.~(\ref{eq:orthogonal}). The ratio appearing in the last line can easily be computed as it involves multinomial distributions, it is equal to
\begin{equation*}
\frac{R_1^{L_1}\ldots R_j^{L_j-1}\ldots R_d^{L_d} \binom{N}{L1,\ldots,L_j-1,\ldots,L_d}}
{R_1^{L_1}\ldots R_j^{L_j}\ldots R_d^{L_d}\binom{N}{L1,\ldots,L_j,\ldots,L_d}} \\
= \frac{1}{R_j} \frac{L_j}N.
\end{equation*}
Inserting this in Eq.~(\ref{eq:prob_ratio}) (and keeping in mind the different definitions of $L_j$) yields the result Eq.~(\ref{eq:single_post}). Averaging this state over measurement outcomes $\bL$ gives
\begin{equation}
\rho_1' = \sum_j R_j \kb{x_j}{x_j}.
\end{equation}

The effect of {\em coarse-grained} measurements $\tilde Q_\el^{(N)}$ on the state of a single molecule can be studied by straightforward modifications of the method outlined above. The results are easily predictable: while the off diagonal  elements $\kb{x_i}{x_j}$ of $\rho_1'$ are completely suppressed when $\sigma =0$, they only get damped by a factor proportional to $Tr\{\tilde Q^{(N-1)}_{\bL^{-x_i}} \tilde Q^{(N-1)}_{\bL^{-x_j}}\}$ when the measurement is coarse. Since $\bL^{-x_i}$ and $\bL^{-x_j}$ are very close to each other on the probability simplex, this decoherence factor is close to unity when the smoothing function $q_\bL(\el)$ is sufficiently wide. 

\subsection{Conditional fidelity}

We will compute, for an ensemble of $N$ two-dimensional molecule initially in state $\beta_1\ket{x_1}+\beta_2\ket{x_2}$, the fidelity between the pre- and the {\em conditional} post-measurement state $\rho_N$ and $\rho_{N|\ell}$ respectively. Again, we will use a single positive number $L$ to denote the type $\bL = (L,1-L)$. The smoothing function is Gaussian $q_L(\ell) = (2\pi\sigma^2)^{-1/2}\exp\{-(L-\ell)^2/2\sigma^2\}$, and we assume that we are in the regime where $\sigma > 1/\sqrt N$. Starting from Eq.~(\ref{eq:ave_rho_post}), we can express the fidelity
\begin{eqnarray*}
F(\rho_N,\rho_{N|\ell}) &=& \bra{\Phi_N}\rho_{N|\ell}\ket{\Psi_N}\\
&=& \sum_{LL'} \sqrt{q_L(\ell)q_{L'}(\ell)} b(L,|\beta_1|^2)b(L',|\beta_1|^2).
\end{eqnarray*}
As $N$ becomes large, we can appeal to the central limit Theorem and approximate a binomial distribution $b(L,|\beta_1|^2)$ by a Gaussian distribution of mean $\mu = |\beta_1|^2$ and variance $\mu(1-\mu)/N$. Moreover, the sum over the discrete values of $L = 0, \frac 1N, \frac 2N, \ldots,1$ can be replaced by an integral over the range [0,1]. These substitutions are accurate within {\em relative} order $O(\frac 1N^2)$ and yield
\begin{eqnarray*}
\int_0^1\int_0^1 \sqrt{\frac{1}{2\pi\sigma^2}} &&e^{-\frac{(L-\ell)^2+(L'-\ell)^2}{2\sigma^2}} \\
&&\times\frac{N}{2\pi \mu(1-\mu)} e^{-N\frac{(L-\mu)^2+(L'-\mu)^2}{2\mu(1-\mu)}} dLdL'.
\end{eqnarray*}
This integral can be performed and yields a complicated expression in terms of ref$(\cdot)$ function, which is not very enlightening. However, under the assumption that $\sigma \gg 1/\sqrt N$, one can derive the lower bound
\begin{eqnarray*}
&&F(\rho_N,\rho_{N|\ell}) \geq \frac{8}
{5\sqrt{2\pi}\sigma} \exp\left\{ \frac{-(\ell-\mu)^2}
{2\sigma^2}\right\} \\
&&\times \left(1-\exp\left\{-\frac{4\sqrt N}{3\sqrt \pi \sigma}\right\} 
-\exp\left\{-\sqrt{2N}\sqrt{\frac{1-p}{p}}\right\}\right)
\end{eqnarray*}
where standard bounds on the error function were used. We see that as long as
\begin{equation}
|\ell-\mu| \leq \Delta^* = \sigma\sqrt{2\ln\frac{1}{c\sigma}}
\label{eq:scaling}
\end{equation}
with $c = 5\sqrt{2\pi}/8$, fidelity will be quasl-exponentially close to one as $N$ increases (``quasi" since the correction is $e^{-\sqrt N}$ instead of strictly $e^{-N}$). Hence, Eq.~(\ref{eq:scaling}) indicates for which measurements outcomes $\ell$ will the conditional post-measurement state $\rho_{N|\ell}$ be quasi-exponentially close to $\rho_N$. 

We can therefore compute the probability that the observed $\ell$ falls in the bound Eq.~(\ref{eq:scaling}), or in other words, the probability that the measurement induces a negligible error to the state of the sample. The probability of the measurement outcome $\tilde Q_\ell^{(N)}$ is given by
\begin{equation}
P(\tilde Q_\ell^{(N)}|\rho_N) = 
\sum_L q_L(\ell) b(\bL,\bR)
\end{equation}
following Eq.~(\ref{eq:probability_ell}). Using the same approximations as above, this gives, up to terms of relative order $O(1/N^2)$,
\begin{equation*}
\int_0^1 \sqrt{\frac{1}{2\pi\sigma^2}}e^{-\frac{(L-\ell)^2}{2\sigma^2}}
 \sqrt{\frac{N}{2\pi \mu(1-\mu)}}e^{-\frac{N(L-\mu)^2}{2\mu(1-\mu)}} dL.
\end{equation*}
Finally, we can substitute the integral over $[0,1]$ by an integral over the real line; as the integrant is positive, this will yield an upper bound to $P(\tilde Q_\ell^{(N)}|\rho_N)$, which turns out to be a Gaussian distribution of mean $\mu$ and variance $\sigma^2+\mu(1-\mu)/N$. Given this upper bound, we can ask what is the probability of getting a measurement outcome $\ell$ which is far from the average outcome $\mu$; 
\begin{eqnarray*}
P(|\ell-\mu|>\Delta) &=& 1- \int_{\mu-\Delta}^{\mu+\Delta} P(\tilde Q_\ell^{(N)}|\rho_N) d\ell \\
&\leq& 1-{\mathrm{erf}}\left\{\frac{\Delta}{\sqrt{2(\sigma^2+\mu(1-\mu)/N)}}\right\} \\
&\leq& e^{-\frac{\sqrt{8}\Delta}{\sqrt{5\pi}\sigma}}
\end{eqnarray*}
where the last line follows from $\sigma > 1/\sqrt N$ and standard bounds on erf$(\cdot)$. Inserting $\Delta^*$ of Eq.~(\ref{eq:scaling}) into this bound gives the probability that the measurement disturbs the state of the ensemble by more than an exponentially small amount. Since $\sigma$ can decrease as fast as $1/\sqrt N$, the probability of appreciably disturbing the state of the ensemble thus scales roughly as $\exp\{-\sqrt{\log N}\}$, and goes to zero as $N$ grows.

\end{document}